\documentclass{jjap2}
\title{Cell Dynamics Simulation of Kolmogorov-Johnson-Mehl-Avrami Kinetics of
Phase Transformation}

\author{Masao \textsc{Iwamatsu}\thanks{E-mail address: iwamatsu@ph.ns.musashi-tech.ac.jp}
and Masato \textsc{Nakamura}\thanks{E-mail address: nakamura@ph.ns.musashi-tech.ac.jp}}

\inst{Department of Physics, General Education Center, Musashi Institute of Technology,\\
Setagaya-ku, Tokyo 158-8557, Japan}

\abst{
In this study, we use the cell dynamics method to test
the validity of the Kormogorov-Johnson-Mehl-Avrami (KJMA) theory 
of phase transformation.
This cell dynamics method is similar to the well-known
phase-field model, but it is a more simple and efficient numerical
method for studying
various scenarios of phase transformation in a unified manner.
We find that the cell dynamics method reproduces the time 
evolution of the volume fraction of the transformed phase 
predicted by the KJMA theory.  Specifically, the cell dynamics
simulation reproduces a double-logarithmic linear KJMA plot and 
confirms the integral Avrami exponents $n$ predicted from the KJMA theory.
Our study clearly demonstrates that the cell dynamics approach is
not only useful for studying the pattern formation but also for
simulating the most basic properties of phase transformation.
}
\kword{phase transformation, cell dynamics, KJMA theory}

\begin{document}
\maketitle

\section{Introduction}
\label{sec:level1}
Phase transformation occurs by the nucleation and subsequent growth of
a nucleus in a system where the first-order phase transformation 
takes place.  It has attracted much attention for more
than a half century~\cite{Kolmogorov,Johnson,Avrami1,Avrami2,Avrami3} from a fundamental
point of view as well as from technological interests.  
These include the mechanical properties of metallic 
materials~\cite{Marx}, the recrystallization of deformed
metals~\cite{Doherty}, and the manufacturing of basic thin-film
transistor devices, such as solar cells
and active matrix-addressed flat-panel displays~\cite{Uemoto}.

The nucleation and growth processes are often described
in terms of the old standard theory called the KJMA theory developed 
by Kolmogorov~\cite{Kolmogorov},
Johnson and Mehl~\cite{Johnson}, and Avrami~\cite{Avrami1,Avrami2,Avrami3}.  
According to this theory, the time evolution of the volume fraction 
of a new transformed phase follows the linear KJMA plot with 
the integral Avrami exponent $n$ that is given by the slope.
However, it is recognized that this theory often fails to explain
experimental results~\cite{Price,Erukhimovitch}; neither the KJMA plot 
becomes linear nor the Avrami exponent becomes an integer.  There is 
also some debate about the validity of the assumption used in 
the theory~\cite{Siclen}.  To resolve the discrepancy, a realistic
yet efficient simulation method that could take various factors 
into account is indispensable.

The direct atomic-scale computer simulation of the kinetics of phase 
transformation using the molecular dynamics or Monte Carlo method is still
a difficult task.  Even the most fundamental phenomenon, such as nucleation, is 
not easy to study using these methods. Instead, the problem of phase
transformation has been tackled using a coarse-grained 
Ginzburg-Landau-type model called 
the Cahn-Hilliard~\cite{Cahn2}, Ginzburg-Landau~\cite{Valls} or 
phase-field~\cite{Jou, Castro} model, which requires the solution 
of highly nonlinear partial differential equations.  Since this model requires
the time integration of the partial differential equations, it is not easy to 
simulate the long-time behavior of the dynamics
of phase transformation~\cite{Roger} except for special traveling wave
solutions~\cite{Chan,Iwamatsu}.

A model based on a cellular automaton instead of a partial differential
equation can improve the efficiency of numerical
integration~\cite{Hesselbarth,Marx}.  It has been used to test the KJMA
theory in the recrystallization of metals.  Although the cellular
automaton is sufficiently flexible to implement the various local reactions of 
recrystallization, it does not have
a direct connection to the equilibrium phase diagram.  Therefore, the connection
between the phase diagram and the phase transformation is not so clear compared to
the Ginzburg-Landau-type model mentioned above. 

In this study, instead, we use the cell dynamics method~\cite{Oono,Puri} to study 
the validity of the KJMA theory.  This method is attractive because it has
the merit of cellular automata and is computationally efficient, and yet
it keeps the connection to the phase diagram through the Landau-type
free energy.  The format of this paper
is as follows: in \S 2, we review the cell dynamics method and 
present the necessary modification for studying the nucleation and growth.  
In \S 3, we follow closely the work by Jou and Lusk~\cite{Jou} and test the validity
of the KJMA theory using the cell dynamics method.  We conclude in section 4.

\section{Cell Dynamics Method for Nucleation and Growth}
\label{sec:sec2}

To study the phase transformation, it is customary to study the
partial differential equation called the phase-field model~\cite{Jou,Castro} which
is equivalent to the time-dependent Ginzburg-Landau (TDGL)~\cite{Valls} or 
Cahn-Hilliard model~\cite{Cahn2}:
\begin{equation}
\frac{\partial \psi}{\partial t}=-\frac{\delta \cal F}{\delta \psi},
\label{eq:2-1}
\end{equation}
where $\delta$ denotes the functional differentiation, $\psi$ is 
the {\it nonconserved} order parameter, and $\cal F$ is the
free energy functional. This free energy is usually written as the square-gradient
form
\begin{equation}
{\cal F}[\psi]=\frac{1}{2}\int \left[D(\nabla \psi)^{2}+h(\psi)\right]{\rm d}{\bf r}. 
\label{eq:2-2}
\end{equation}
The local part $h(\psi)$ of the free energy $\cal F$ determines the bulk phase
diagram and the value of the order parameter in equilibrium phases.
The double-well form was frequently used for $h(\psi)$ to express 
the two-phase coexistence and study the phase transformation 
between these two phases.

This TDGL equation (\ref{eq:2-1}) for the nonconserved order parameter
$\psi$ was loosely transformed into a space-time discrete cell dynamics
equation by Puri and Oono~\cite{Puri} following a similar transformation
of the kinetic equation for the conserved order parameter called
the Cahn-Hilliard-Cook equation~\cite{Oono}.  In their cell dynamics method, the partial
differential equation (\ref{eq:2-1}) is replaced by a finite difference
equation in space and time in the form
\begin{equation}
\psi(t+1,n)=F[\psi(t,n)],
\label{eq:2-4}
\end{equation}
where the time $t$ is discrete and an integer, and the space is also discrete and
is expressed by the integral site index $n$.  The mapping $F$ is given by
\begin{equation}
F[\psi(t,n)]=-f(\psi(t,n))+\left[\ll\psi(t,n)\gg-\psi(t,n)\right],
\label{eq:2-5}
\end{equation}
where $f(\psi)=dh(\psi)/d\psi$, and the definition of $\ll\cdots\gg$ for 
a two-dimensional square grid is
given by~\cite{Oono,Puri,Teixeira1}
\begin{equation}
\ll\psi(t,n)\gg=\frac{1}{6}\sum_{i=\mbox{nn}}\psi(t,i)
+\frac{1}{12}\sum_{i=\mbox{nnn}}\psi(t,i),
\label{eq:2-6}
\end{equation}
where ``nn'' denotes nearest neighbors and ``nnn'' next-nearest neighbors.  
An improved form of this mapping for a three-dimensional case was also
obtained~\cite{Teixeira1}.

Oono and Puri~\cite{Oono,Puri} further approximated the derivative
of the local free energy $f(\psi)$ called the ``map function'' in the $\tanh$ form
\begin{equation}
f(\psi)=\frac{dh}{d\psi}\simeq \psi-A \tanh \psi,
\label{eq:2-7}
\end{equation}
with $A=1.3$ that corresponds to the free energy~\cite{Cakrabarti}
\begin{equation}
h(\psi)=-A\ln\left(\cosh\psi\right)+\frac{1}{2}\psi^{2}
\label{eq:2-8}
\end{equation}
and is the lowest order ($O(\psi^{2})$) approximation to the double well form
of the free energy
\begin{equation}
h(\psi)=-\frac{1}{2}\psi^{2}+\frac{1}{4}\psi^{4}
\label{eq:2-8x}
\end{equation}
when $A=1.5$.  They~\cite{Oono,Puri} used this simplification 
since this cell dynamics system is invented not to {\it simulate} the
mathematical TDGL partial differential equation but to {\it simulate} and
{\it describe} the behavior of nature directly.  
Later, Chakrabarti and Brown~\cite{Cakrabarti} discussed that this simplification
is justified since the detailed form of the double-well potential $h(\psi)$ 
is irrelevant to the long-time dynamics and the scaling exponent.

Subsequently, however, several authors used the map function $f(\psi)$ directly
obtained from the free energy $h(\psi)$ as it is~\cite{Qi,Ren1} and found that
the cell dynamics equation (\ref{eq:2-4}) is still tractable numerically. 
Ren and Hamley~\cite{Ren1} argued that one can easily include the 
effect of the asymmetry of the free energy and the asymmetric 
characteristic of two phases using the original form of the
free energy function $f(\psi)$.

Despite the popularity of this cell dynamics method in the soft-condensed
matter community as a simulator of pattern formation due to various
factors, it has not yet been used to study the most
fundamental problem of phase transformation by nucleation and growth.  
In the next section,
we use a parameterized free energy function, 
study the kinetics of phase 
transformation, and test the validity of the KJMA theory.

\section{Numerical Results}
\subsection{Two-dimensional growth of single domain}
To simulate the growing stable phase after the
nucleation, we have to prepare the system in a state where one phase
is metastable and has the higher free energy than the other stable phase.  
The free energy difference 
between the stable and metastable phases is determined from the
supersaturation in liquid condensation and from the
undercooling in crystal nucleation.  Microscopically, this
free energy difference is necessary for the nucleus to overcome the
additional curvature effect caused by the interfacial tension and 
to continue to grow~\cite{Iwamatsu2}.  

\begin{figure}[htbp]
\begin{center}
\includegraphics[width=0.6\linewidth]{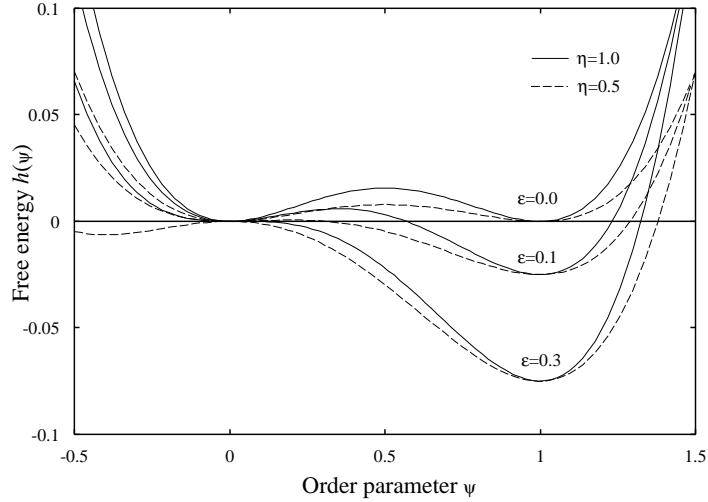}
\end{center}
\caption{
Model double-well free energy (\ref{eq:3-1}) that can realize
two-phase coexistence when $\epsilon=0$.  The parameter $\epsilon$ determines the
free energy difference $\Delta h$, and the parameters $\eta$ and 
$\epsilon$ determine
the free energy barrier $\Delta E$.
}
\label{fig:1}
\end{figure}

To study the growth of the stable phase using the cell dynamics 
method, we consider 
the time-dependent Ginzburg-Landau (TDGL) equation (\ref{eq:2-1}) with the 
square gradient free energy functional (\ref{eq:2-2}).
The local part of the free energy $h(\psi)$ we used is~\cite{Jou}
\begin{equation}
h(\psi) = \frac{1}{4}\eta\psi^{2}(1-\psi)^{2} + \frac{3}{2}\epsilon\left(\frac{\psi^{3}}{3}-\frac{\psi^{2}}{2}\right).
\label{eq:3-1}
\end{equation}
This free energy is shown in Fig. \ref{fig:1}, where one phase at $\psi=0$ 
is metastable while another phase at $\psi=1$ is stable.
The free energy difference $\Delta h$ between the stable phase
at $\psi=1$ and the metastable phase at $\psi=0$ is solely 
determined from the parameter $\epsilon$:
\begin{equation}
\Delta h = h(\psi=0)-h(\psi=1)=\frac{\epsilon}{4}.
\label{eq:3-1x}
\end{equation}
Therefore, $\epsilon$ represents the supersaturation or the
undercooling.

The metastable phase at $\psi=0$ becomes unstable 
when $\eta=3\epsilon$, which defines the spinodal.
The height $\Delta E$ of the free energy barrier at $\psi=(\eta-3\epsilon)/2\eta$ can 
be tuned by the parameters $\eta$ and $\epsilon$: 
\begin{equation}
\Delta E=h\left(\psi=\frac{\eta-3\epsilon}{2\eta}\right)-h(\psi=0)
=\frac{\eta^{4}-4\eta^{3}\epsilon+27\epsilon^{4}}{32\eta^{3}},
\label{eq:3-1y}
\end{equation}
which vanishes when $\eta=3\epsilon$ at the spinodal.

The steady-state analytical solution of the TDGL with a constant 
interfacial velocity has been
obtained in one dimension by Chan~\cite{Chan} when the 
free energy $h(\psi)$ is written in the
quadratic form such as in eq. (\ref{eq:3-1}).   Using Chan's formula, the interfacial
velocity $v$ of our TDGL model ({\ref{eq:2-1}), (\ref{eq:2-2}) with the 
free energy (\ref{eq:3-1}) is given by
\begin{equation}
v = \sqrt{\frac{D}{2\eta}}3\epsilon = \sqrt{\frac{D}{2\eta}}12\Delta h.
\label{eq:3-2}
\end{equation}    
Chan~\cite{Chan} further suggested that if the interfacial
width is small, the interfacial velocity of a two-dimensional 
circular or three-dimensional spherical growing nucleus is
asymptotically given by eq. (\ref{eq:3-2}) of the one-dimensional model.  
The larger the
free energy difference $\epsilon$ and the lower the free energy
barrier $\eta$ are, the higher the front velocity $v$ becomes.

The critical radius $R_{c}$ of a two-dimensional circular nucleus is
also given analytically~\cite{Chan,Jou}:
\begin{equation}
R_{c}=\frac{D}{v}=\frac{\sqrt{2\eta D}}{3\epsilon}.
\label{eq:3-3}
\end{equation}
In the metastable phase, the circular nucleus of the stable phase with 
a radius ($R$) smaller than $R_{c}$ shrinks, while that with 
a radius larger than $R_{c}$ grows and its front velocity 
approaches eq. (\ref{eq:3-2}).  Again, the larger the
free energy difference $\epsilon$ and the lower the free energy
barrier $\eta$ are, the smaller the critical radius $R_{c}$ is.

We implemented the above free energy (\ref{eq:3-1}) into the
cell dynamics code written by Mathematica TM~\cite{Wolfram} 
for the animation of 
spinodal decomposition developed by Gaylord and Nishidate~\cite{Gaylord},
and simulated the growth of a single circular nucleus of a stable phase.

Initially, we prepared a small circular nucleus of
a stable phase within a metastable phase and simulated
the growth of that nucleus.  The system size is 100$\times$100 cells, 
$D$ in eq. (\ref{eq:2-2}) is $D=0.5$, and
the periodic boundary condition is imposed.
The initial order
parameter $\psi$ is randomly chosen from $0.9\leq \psi \leq 1.1$ for 
the circular nucleus of the stable phase and from $-0.1\leq \psi \leq 0.1$ for 
the metastable environment.  The diameter of the initial nucleus is fixed
at $d=11$.  Therefore, the initial nucleus occupies a part of 
11$\times$11 cells.  
The effective area of the stable phase is computed by
counting the number of cells with the order parameter $\psi>1/2$.

\begin{figure}[htbp]
\begin{center}
\includegraphics[width=0.6\linewidth]{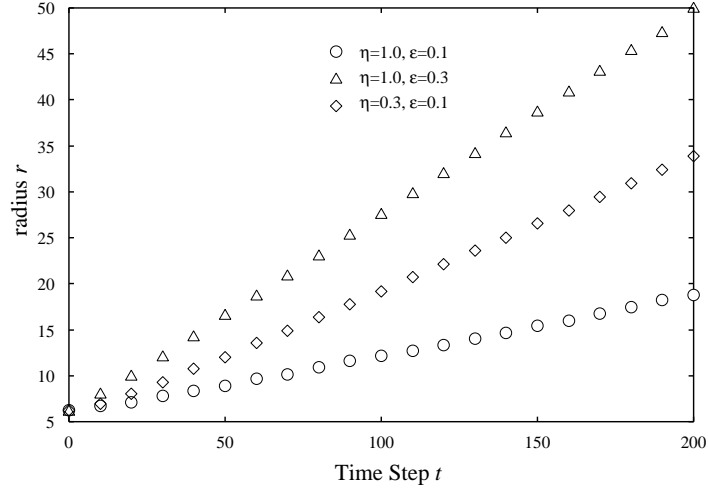}
\end{center}
\caption{
Evolution of effective radius of area of stable 
phase plotted as a function of time when $\eta=1$ and 
$\epsilon=0.1$ (open circles), $\eta=1.0$ and $\epsilon=0.3$ 
(open triangles), and $\eta=0.3$ and $\epsilon=0.1$ (open diamonds).
The effective radius shows a linear time dependence, indicating a
constant interfacial velocity.
}
\label{fig:2}
\end{figure}

Figure \ref{fig:2} shows the effective radius of the circular nucleus
of the stable phase calculated from the effective area 
of the nucleus as a function of
time step $t$.  The nearly linear growth of the 
nucleus of the stable phase is clearly visible, which indicates a constant 
front velocity for the stable-metastable interface predicted from the analytical
solution for the TDGL~\cite{Chan}. The velocities $v$ estimated from 
Fig. \ref{fig:2} and predicted from eq. (\ref{eq:3-2}) are
compared in Table ~\ref{tab:1} together with the critical
radius $R_{c}$ calculated from eq. (\ref{eq:3-3}).  
It can be seen that eq. (\ref{eq:3-2}) 
correctly predicts the general trend of the front velocity $v$ when 
the two parameters $\eta$ and $\epsilon$ are
altered.  Since 
the cell dynamics method {\it does not} solve 
the TDGL directly, the discrepancy between cell dynamics simulation 
and theoretical prediction (\ref{eq:3-2}) by a factor of roughly 2 
is not very significant.

\begin{table}[htbp]
\caption{Comparison of front velocities $v$ estimated by cell dynamics
simulation and prediction using eq. (\ref{eq:3-2}) for various
potential parameters $\eta$ and $\epsilon$.  The corresponding critical
radii $R_{c}$ are also tabulated.}
\label{tab:1}
\begin{tabular}{ccccccc}
\hline
$\eta$            &  1     &  1   &   0.7 &  0.5  &   0.4 & 0.3   \\  
$\epsilon$        &  0.1   & 0.3  &   0.1 &   0.1  &  0.1 &  0.1   \\
\hline
$v$ (simulation)        &  0.063  & 0.22 &   0.082  &  0.10 & 0.12 & 0.14  \\
$v$ [eq. (\ref{eq:3-2})] &  0.15  & 0.45 &   0.18  &  0.21 &  0.24  &  0.27  \\
$R_{c}$ [eq. (\ref{eq:3-3})] & 3.3 & 1.1 & 2.8 & 2.4 & 2.1 & 1.8 \\
\hline
\end{tabular}
\end{table}

From the above comparison of the values estimated by cell dynamics 
simulation and theoretical
prediction using the steady-state solution of the TDGL for a two-phase 
system, we consider that
this cell dynamics method is effective for studying the phase 
transformation by the growth of the multiple nucleus of the stable phase.

\subsection{KJMA kinetics by cell dynamics simulation}

\subsubsection{Site-saturation nucleation}

In site-saturation
nucleation, a fixed number of nuclei are prepared initially, and
subsequent growth is monitored.  The KJMA theory gives an analytical 
expression for the volume fraction $f$ of the stable phase as a
function of time $t$.  In two dimensions, the formula leads to~\cite{Jou}
\begin{equation}
f = 1 - \exp\left(-\pi n_{0}v^{2}\left(t+t_{0}\right)^{2}\right),
\label{eq:4-1}
\end{equation}
where $n_{0}$ is the number density (number per unit area) of 
the randomly distributed initial nuclei.  $v$ is the
growth rate of the radius of each nucleus discussed in the 
previous section. $t_{0}$ is
the origin of time which can hopefully take the incubation time of
nucleation into account~\cite{Castro}.

From eq. (\ref{eq:4-1}), we have
\begin{equation}
\log \left(-\ln (1-f)\right)=2\log \left(t+t_0 \right) + \mbox{constant}.
\label{eq:4-2}
\end{equation}
Therefore, the KJMA theory predicts that a double logarithms
$\log \left(-\ln (1-f)\right)$ versus $\log\left(t+t_{0}\right)$ is
a straight line that is known as the KJMA plot with 
the integral tangent $n=2$, which is called the ``Avrami exponent''.

\begin{figure}[hptb]
\begin{center}
\includegraphics[width=0.6\linewidth]{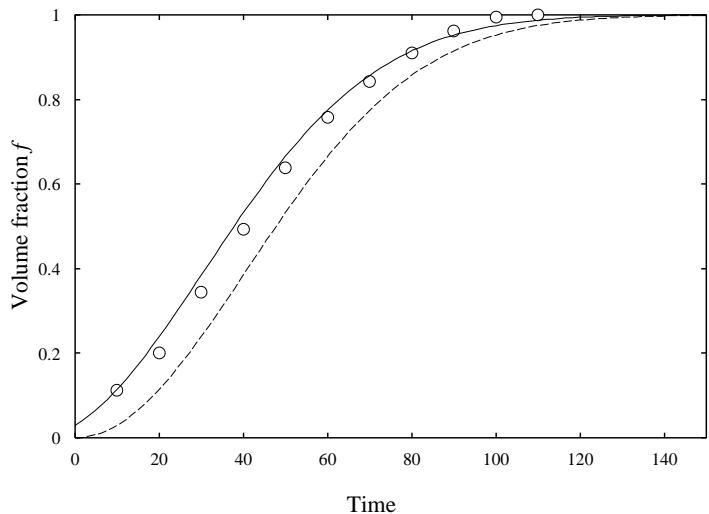}
\end{center}
\caption{
Evolution of volume fraction $f$ calculated by 
cell dynamics simulation for site-saturation nucleation
as a function of time (open circles) when $\eta=1.0$ and $\epsilon=0.3$.  
The broken line denotes the theoretical prediction (\ref{eq:4-1})
with $t_0=0$, while the solid line denotes that with $t_0=10$.
A better agreement between the simulation and theoretical results
is attained when the incubation time $t_0=10$ is taken into account.  
}
\label{fig:3a}
\end{figure}

We have simulated the site-saturation nucleation using
the cell dynamics method.  Now a finite number of nuclei
of the stable phase is distributed over the area we considered.  
The initial nuclei are circular and have the diameter
$d=8$, which is larger than the critical 
radius $R_{c}$ in Table ~\ref{tab:1}.  Then, the evolution of the transformed
volume is monitored as a function of time. Again, we have
considered the 100$\times$100 system and introduced
a finite number ($N_{0}=20$) of nuclei as the initial condition.
Therefore, the number density of the initial nucleus is $n_{0}=20/10000=0.002$.

The time evolution of the transformed volume $f$ is plotted as
a function of time $t$ in Fig. \ref{fig:3a}.  When the effect
of the incubation time with $t_0=10$ is included, a better agreement 
between the simulation and theoretical results
is attained.  This incubation time $t_{0}=10$ is the time
necessary for a infinitely small nucleus to become a larger
nucleus with the diameter $d=8$ in our simulation, and is 
estimated by fitting
the theoretical curve (\ref{eq:4-1}) to the simulation data.
Since an infinitely small nucleus cannot grow because it is
smaller than the critical nucleus, we use the terminology
``incubation time''
to indicate both the time necessary for 
a critial nucleus to appear and the time necessary for it
to grow to be a larger nucleus.

The KJMA plot
of the double logarithm of the volume fraction $f$ is shown as a
function of $\log t$ in Fig.~\ref{fig:3b}, where we ignore the
effect of the incubation time and set $t_{0}=0$. The time 
evolutions for several
combinations of the potential parameters $\eta$ and $\epsilon$
are shown.  They do not fit the expected straight lines.

\begin{figure}[hptb]
\begin{center}
\includegraphics[width=0.6\linewidth]{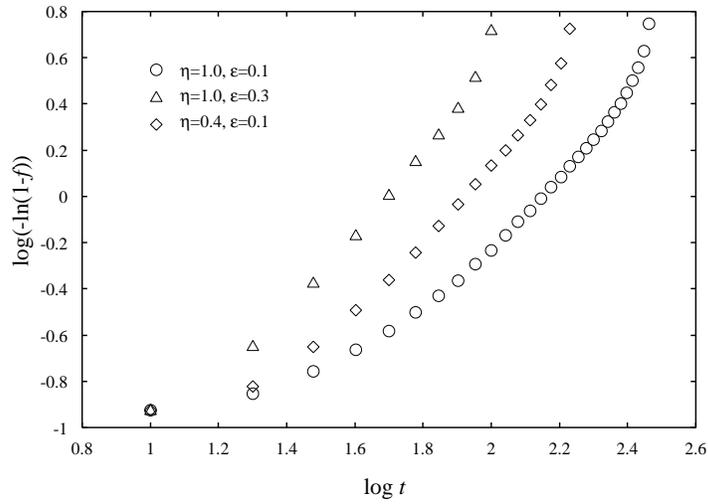}
\end{center}
\caption{
Evolution of volume fraction $f$ calculated by 
cell dynamics simulation for site-saturation nucleation
as a function of time when $t_0=0$.  
The double logarithm
$\log \left(-\ln (1-f)\right)$ is plotted as a function 
of the logarithm of time $\log t$.  All data do not
fit the straight lines predicted from the KJMA theory (\ref{eq:4-2}). 
}
\label{fig:3b}
\end{figure}

Figure \ref{fig:3c} shows the KJMA plot when the incubation
time $t_{0}$ is considered.  Again, the incubation time $t_{0}$ is
estimated by fitting the theoretical curve ({\ref{eq:4-1}) to
the simulation data.  Now, the time evolutions for several
combinations of the potential parameters $\eta$ and $\epsilon$
all fit the straight lines with almost the same Avrami exponent
$n\simeq 2$, which is very close to the theoretical prediction, 
as shown in Table ~\ref{tab:2}. 
The results in Figs. \ref{fig:3b} and \ref{fig:3c} clearly suggest that 
the incubation time $t_{0}$ should be carefully taken into account 
to deduce the Avrami exponent $n$ when we analyze experimental as well as 
simulation data. 

\begin{figure}[hptb]
\begin{center}
\includegraphics[width=0.6\linewidth]{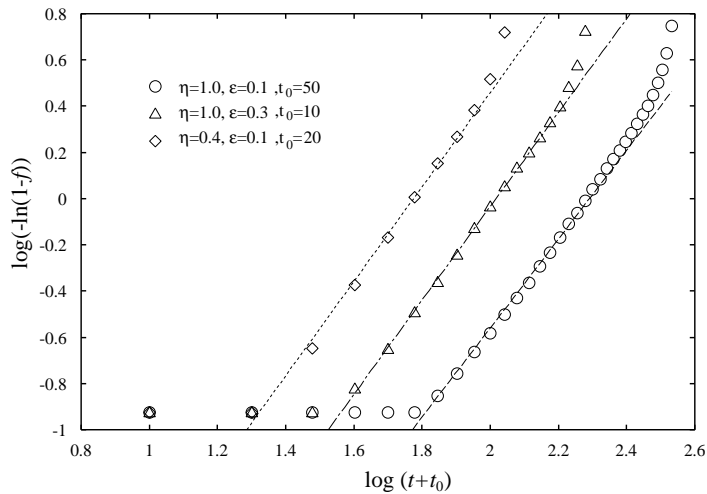}
\end{center}
\caption{
Same as Fig.~\ref{fig:3b} when incubation time
$t_{0}$ is taken into consideration.  All data follows
an almost straight line and confirm
the prediction of the KJMA theory given by eq. (\ref{eq:4-2}) with
all the Avrami exponents $n$ deduced from the straight lines
in the figure being close to the theoretically
predicted $n=2$. 
}
\label{fig:3c}
\end{figure}

\begin{table}[htbp]
\caption{Avrami exponent $n$ for site-saturation nucleation 
estimated by cell dynamics simulation for various
potential parameters $\eta$ and $\epsilon$.  The value 
theoretically predicted from the KJMA theory is $n=2$.}
\label{tab:2}
\begin{tabular}{cccc}
\hline
$\eta$            &  1     &  1   &   0.4 \\  
$\epsilon$        &  0.1   & 0.3  &   0.1 \\
\hline
$n$ (simulation)        &  1.92  & 2.04    &   2.03  \\
\hline
\end{tabular}
\end{table}

Figure \ref{fig:4} shows the evolution of the morphology of
the two-dimensional system for the site-saturation
nucleation when $\eta=0.4$ and $\epsilon=0.1$. 
We observe the almost isotropic
growth of every nucleus of the stable phase.  
At the time step $\sim$100, almost all cells 
are transformed into the stable phase.

\begin{figure}[htbp]
\begin{center}
\includegraphics[width=0.8\linewidth]{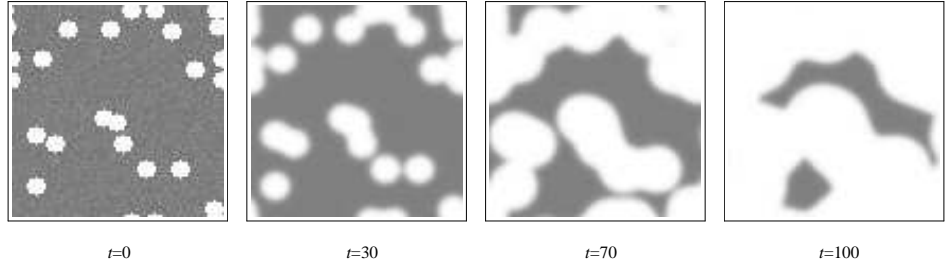}
\end{center}
\caption{
Typical evolution pattern of site-saturation nucleation
calculated by cell dynamics simulation
when $\eta=0.4$ and $\epsilon=0.1$.
The white area indicates the stable phase.
}
\label{fig:4}
\end{figure}

\subsubsection{Continuous nucleation}
In the continuous
nucleation, a new nuclear embryo is continuously introduced.  
The KJMA theory of continuous nucleation gives the analytical expression for
the volume fraction $f$ of the growing stable phase. In two dimensions,
it leads to~\cite{Jou}
\begin{equation}
f = 1 - \exp\left(-\frac{\pi\dot{n}v^{2}}{3}\left(t+t_0\right)^{3}\right),
\label{eq:4-3}
\end{equation}
where $\dot{n}$ is the steady nucleation rate per unit area and $v$ is the
growth rate of the single nucleus discussed in the previous section.

From eq. (\ref{eq:4-3}), we have
\begin{equation}
\log \left(-\ln (1-f)\right)=3\log\left(t+t_{0}\right) + \mbox{constant}.
\label{eq:4-4}
\end{equation}
Therefore, a double logarithmic KJMA plot should give
the ``Avrami exponent'' $n=3$ instead of $n=2$ of the site-saturation
nucleation.

We have also simulated the continuous nucleation using
the cell dynamics method.  
In our simulation, a constant nucleation rate $\dot{n}$ is
achieved by introducing a new nucleus every $1/\dot{n}$ time
step (nucleation time).  At each nucleation time step, a position 
within the two-dimensional area is randomly selected.  If the
position is already occupied by the stable phase, no new nucleus
is placed.  If the position is not occupied by the stable
phase, a new nucleus is placed and allowed to grow there. 
In this simulation, we have
used a larger 200$\times$200 system to avoid the finite-size
effect as much as possible.  The steady nucleation 
rate $\dot{n}=0.1/40000$ is used.  Therefore,
a single nucleus is produced at every 10 time steps in the area 200$\times$200. 

The time evolution of the transformed volume $f$ is plotted as
a function of time $t$ in Fig. \ref{fig:5} as the double logarithmic
KJMA plot. The time evolutions for several
combinations of the potential parameters $\eta$ and $\epsilon$
show almost the same straight line with the Avrami exponent
very close to the theoretically predicted $n=3$, as shown in Table ~\ref{tab:3}.

\begin{figure}[htbp]
\begin{center}
\includegraphics[width=0.6\linewidth]{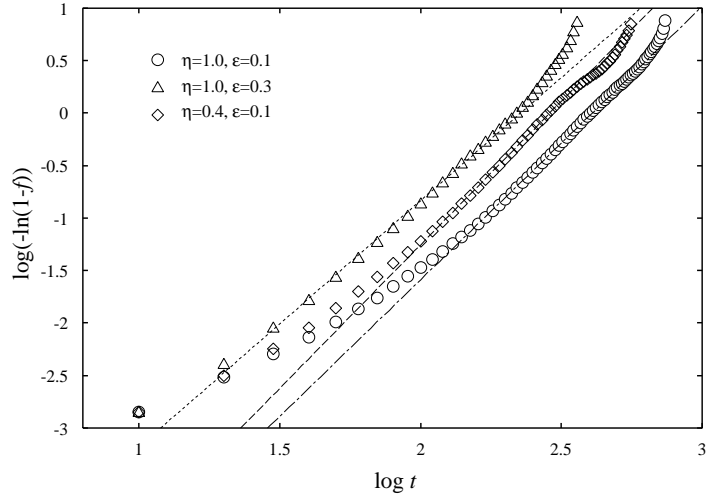}
\end{center}
\caption{
Evolution of volume fraction $f$ calculated by 
cell dynamics simulation for continuous nucleation
as a function of time.  The double logarithmic
KJMA plot is used. All curves fit the straight lines
predicted from the KJMA theory (\ref{eq:4-4}).
The Avrami exponents deduced from the straight lines in the
figure are all close to the theoretically
predicted $n=3$, as summarized in Table ~\ref{tab:3}. 
}
\label{fig:5}
\end{figure}

\begin{table}[htbp]
\caption{Avrami exponent $n$ for continuous nucleation 
estimated by cell dynamics simulation for various
potential parameters $\eta$ and $\epsilon$.  The value 
theoretically predicted from the 
KJMA theory is $n=3$.}
\label{tab:3}
\begin{tabular}{cccccc}
\hline
$\eta$            &  1.0     &  1.0   &   0.4 \\  
$\epsilon$        &  0.3     &  0.1   &   0.1 \\
\hline
$n$ (simulation)        &  2.35   &  2.60   &   2.73\\
\hline
\end{tabular}
\end{table}

Figure \ref{fig:6} shows the evolution of the morphology of
the two-dimensional systems for the continuous nucleation when $\eta=0.4$ 
and $\epsilon=0.1$.  Because a nucleus is continuously
produced, almost all cells are occupied at the later stage, and
then the production of a nucleus stops.  The situation becomes
closer to the site-saturation nucleation.  The Avrami exponent $n$ 
smaller than the theoretically predicted $n=3$ for the continuous
nucleation but closer to the site saturation nucleation $n=2$ is
expected.  This finite-size effect is 
one of the reasons why the Avrami exponent estimated from the simulation
data in Table ~\ref{tab:3} is always smaller than the theoretically
predicted $n=3$.

There is also a problem of incubation time
in continuous nucleation.  In our simulation, we have
introduced a fairly large nucleus, which is sufficiently large
to grow continuously.  Thus, the same problem of
time origin or incubation time $t_{0}$ as that for
the site-saturation nucleation could occur.
Since a nucleus is continuously produced, we could not
incorporate the effect of incubation time in a reasonable
manner in our analysis.

\begin{figure}[htbp]
\begin{center}
\includegraphics[width=0.8\linewidth]{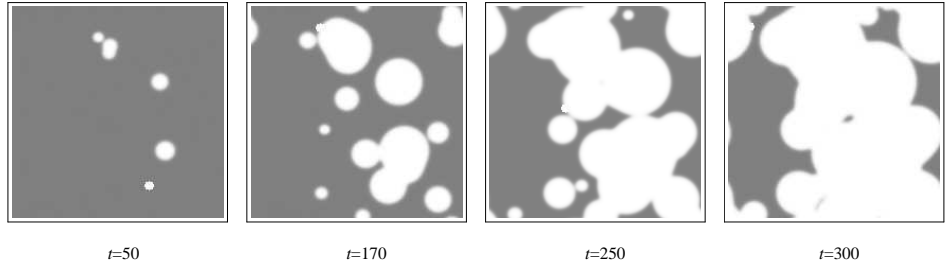}
\end{center}
\caption{
Typical evolution pattern of continuous nucleation simulation
when $\eta=0.4$ and $\epsilon=0.1$.
}
\label{fig:6}
\end{figure}

In continuous nucleation, we found that the KJMA theory correctly describes the
overall behavior of the time dependence of the transformed volume fraction $f$.
However, there is a small inflection in the slope of the KJMA plot at around
$\log(-\ln (1-f))\approx 0.0$, which can be explained by the impingement
in which the growing circular grains collide with each other.  
The volume fraction $f$ when the impingement starts to occur is 
roughly estimated to be $f=\pi r^{2}/4r^{2}\approx 0.79$
using the ratio of the area of a circular
grain with the radius $r$ to that of a square with the side length $2r$.
Thus, the inflection of the 
KJMA straight line is expected at
\begin{equation}
\log(-\ln (1-f))=\log(-\ln (0.21))\approx 0.19,
\end{equation}
which is very close to the point where the inflection is actually observed
in Fig.~\ref{fig:5}.  The same effect was discussed by Jou and
Lusk~\cite{Jou}.  The other factors affecting the
Avrami exponent are discussed in the next subsection.

\subsection{Discussion}

Experimentally, a reasonably linear KJMA behavior was observed in
the recrystallization of some metals and in the crystallization of
metallic glasses~\cite{Price}.  However, a considerable
variation in Avrami exponent $n$ defined by
\begin{equation}
\log \left(-\ln (1-f)\right)=n\log\left( t+t_{0}\right) + \mbox{constant}
\label{eq:5-1}
\end{equation}
was observed from the electrical resistivity and
differential scanning calorimetry (DSC) data.

Following the argument of Christian~\cite{Christian},
Price~\cite{Price} proposed a formula for the Avrami
exponent
\begin{equation}
n=a+b(1-q),
\label{eq:5-2}
\end{equation}
where $a$ is the nucleation component; $a=0$ for the site saturation
and $a=1$ for the continuous nucleation. $b$ defines the dimensionality
of the growth ($b=3$ for a three-dimensional problem and $b=2$ for our
two-dimensional problem).  The exponent $q$ includes contributions from
various types of power-law reaction. 

For example, in our cell dynamics model based on the Landau-type free 
energy (\ref{eq:3-1}), the driving force of phase transformation
comes from the undercooling defined by eq. (\ref{eq:3-1x}) that leads to
the linear time-dependent growth of a circular nucleus with the constant
front velocity $v$ given by eq. (\ref{eq:3-2}).  However, as
more materials are transformed into the stable phase, the driving force
decreases somehow and the front velocity $v$ is expected
to decelerate from eq. (\ref{eq:3-2}).  Thus, it is reasonable to assume 
a power-law decay of the front velocity
\begin{equation}
v \propto t^{-q},
\label{eq:5-3}
\end{equation}
which exactly gives eq. (\ref{eq:5-2}).  Hence, the
Avrami exponent becomes smaller than the KJMA predicted
$n=a+b$, as observed in our numerical simulations for the continuous nucleation.

There are many other factors affecting the exponent $n$. Possible
reasons for the nonideal exponent and even the nonlinear growth in
the KJMA plot include the nonrandomness of the nucleation site and the 
preferential nucleation, for example, at
the grain boundary~\cite{Cahn3}, the effect of the time dependence
of the nucleation rate and so forth.  The net result of these effects is
a negative deviation from the KJMA linear plot~\cite{Price}, which leads to
again the {\it smaller} exponent $n$ in accordance to the many
experimental results and our simulation. In our cell dynamics method, 
these effects
can be easily included by changing the probability of selecting
the nucleation rate from cell to cell.  Hesselbarth and
G\"obel~\cite{Hesselbarth} have included such effects in their
cellular automaton and could successfully explain the
deviation of experimental data from KJMA predicted data.

There is also a problem of two-stage crystallization~\cite{Price,Burton}.  
In some alloys, the KJMA
plot shows an inflection in which the exponent $n$ changes 
markedly from
a large value at an early stage to a small value at a later stage.
However, our simulation data shown in Fig. \ref{fig:5} shows
the opposite trend; the exponent $n$ is large at the later stage.
This phenomena is explained by assuming that the early
stage corresponds to the continuous nucleation and that the later
stage corresponds to the site saturation because of the exhaustion
of the nucleation site~\cite{Burton}.  Recent theoretical model calculation
supports this two-stage nucleation model~\cite{Rollett,Castro2}.

Our cell dynamics method could easily incorporate such a two-stage
transformation by assuming that the continuous nucleation terminates
at a certain stage.  Then, the growth process is continuous 
nucleation with the exponent $n=3$
in the early stage, but it becomes site saturation with the exponent
$n=2$ in the later stage.  In our cell dynamics 
method, it is not necessary to assume a discrete 
lattice~\cite{Rollett,Castro2} and is easier to incorporate various
modifications to KJMA kinetics.  Using
our cell dynamics method, a more quantitative study is 
feasible in the future.

\section{Conclusion}
\label{sec:sec4}

In this study, we used a cell dynamics method to 
test the validity of the Kolmogorov-Johnson-Mehl-Avrami (KJMA) kinetic 
theory of phase transformation.  First, we used 
this method to study the growth of a single
circular nucleus and found that the nucleus grows with a constant
front velocity in accordance to the analytical solution~\cite{Chan}.
Next, we used the cell dynamics method to simulate the growth of
an ensemble of nuclei under the conditions of both the site saturation and 
continuous nucleation.  We found a nearly linear behavior
of the KJMA plot with the Avrami exponent close to the KJMA predicted value.
Finally, we suggested several extensions of the cell dynamics
method to study various contributions that may lead to the 
nonlinear KJMA behavior or nonideal Avrami exponent.

The results obtained in this study are summarized as follows:
\begin{itemize}
\item The cell dynamics method with a realistic free energy can succesfully
simulate the steady growth of a single nucleus and confirm the 
prediction of Chan~\cite{Chan} based on the time-dependent 
Ginzburg-Landau (TDGL) equation.
\item It can also simulate the growth of multiple nuclei and confirm the time
evolution of the volume fraction of the transformed material predicted from 
the KJMA kinetic theory~\cite{Kolmogorov} and numerical simulation using TDGL~\cite{Jou}.
\item Therefore, the cell dynamics method can be used to simulate more complex scenarios of
nucleation and growth.
\item Our simulation indicates that the incubation time should be carefully
taken into account when we deduce the Avrami exponent 
from the experimental and simulation data.
\end{itemize}

The cell dynamics method is similar to the time-dependent Ginzbug-Landau
or Cahn-Hilliard model based on the free energy functional.  In contrast 
to the conventional cellular automaton approach
to the phase transformation~\cite{Hesselbarth,Marx,Rollett,Castro2}, 
no phenomenological energy that induces phase transformation
is necessary.  Therefore, the cell dynamics method is 
numerically efficient as a cellular automaton, yet it keeps the
direct connection to the equilibrium phase diagram.  This cell dynamics
method can be used
to test various scenarios of nucleation and growth in a unified manner.







\end{document}